\documentclass[12pt]{article}
\usepackage{rotating}
\usepackage{pbox}
\usepackage{amsfonts}
\usepackage{amsmath}
\usepackage{amssymb}
\newcommand{\be}{\begin{equation}}
\newcommand{\ba}{\begin{eqnarray}}
\newcommand{\ee}{\end{equation}}
\newcommand{\ea}{\end{eqnarray}}

\newcommand{\cosec} { {\rm cosec}}

\begin{document}

\title{A class of exactly solvable rationally extended non-central potentials
in Two and Three Dimensions}

\author{Nisha Kumari$^{a}$\footnote{e-mail address: nishaism0086@gmail.com (N.K)}, Rajesh Kumar Yadav$^{b}$\footnote {e-mail address: rajeshastrophysics@gmail.com (R.K.Y)}, Avinash Khare$^{c}$\footnote{e-mail: khare@physics.unipune.ac.in (A.K)} and \\ Bhabani Prasad Mandal$^{a}$\footnote{e-mail address: bhabani.mandal@gmail.com (B.P.M).}}
 \maketitle
 {$~^a$Department of Physics, Banaras Hindu University, Varanasi-221005, India.\\
 $~^b$Department of Physics, S. P. College, S. K. M. University, Dumka-814101, India.\\
 $~^c$Department of Physics, Savitribai Phule Pune University, Pune-411007, India.\\
}

\begin{abstract}
We start from a seven parameters (six continuous and one discrete) family of 
non-central exactly solvable 
potential in three dimensions and construct a wide class of ten parameters 
(six continuous and four 
discrete) family of rationally extended exactly solvable 
 non-central real as well as  $PT$ symmetric complex potentials. 
The energy eigenvalues and the eigenfunctions of these extended non-central 
potentials are obtained explicitly and it is shown that the wave eigenfunctions 
of these potentials are either associated with the exceptional orthogonal 
polynomials (EOPs) or some type of new polynomials which can be further 
re-expressed in terms of the corresponding classical orthogonal polynomials. 
Similarly, we also construct a wide class of rationally extended exactly 
solvable non-central real as well as complex PT-invariant potentials in 
two dimensions.
\end{abstract}
\section{Introduction}
In non-relativistic quantum mechanics the exactly solvable (ES) problems play 
an important role in the understanding of different quantum mechanical 
systems associated with any branch of theoretical 
physics. For many of the quantum mechanical systems, whose exact solutions 
are unknown, these ES potentials are generally 
considered as a starting potential to get their approximate eigenspectrum. 
Most of the ES potentials are either one dimensional   
 or are central potentials which are essentially one dimensional on the half 
line. 

There are only few examples of exactly solvable non-central potentials such as 
anisotropic harmonic oscillator whose solutions are well known, see for example
\cite{fluge}. In a comprehensive study, Khare and Bhaduri \cite{kb} have 
discussed a  number of exactly solvable non-central potentials
 by considering the Schr\"odinger equation in 
three dimensional spherical polar and two dimensional polar co-ordinates. The classical orthogonal polynomials (such as 
Hermite, Laguerre and Jacobi polynomial etc) 
are playing a fundamental role in the construction of the bound state 
solutions to these ES potentials. The solution 
of most of the ES non-central potentials is connected with the above well known
 orthogonal polynomials. It has been observed that the solution
of all the non-central potentials obtained in Ref. \cite{kb} is associated 
either with the classical Laguerre or Jacobi orthogonal polynomials. 

After the recent discovery of the two new families of orthogonal polynomials
 namely the exceptional $X_m$ Laguerre and 
$X_m$ Jacobi orthogonal polynomials \cite{dnr1, xm1}, a number of new exactly 
solvable potentials have been discovered 
\cite{que,bqr,os,hos,hs,qu,cq_12,yg1,yg2,dim}. In most of these cases,
these new potentials are the rational extension of the corresponding 
conventional potentials \cite{cks, levai89}. Various properties of these 
new extended potentials have also been studied by different groups 
\cite{pdm,nfold2,fplank,codext1,qscat,scatpt,gtextd,para_sym,bpm5,nrab}. 
It is then natural to consider the rational extension of 
the conventional non-central potentials discussed in Ref. \cite {kb} and 
discover new exactly solvable non-central potentials whose solutions are in
terms of the rational extension of the Jacobi or Laguerre polynomials.
The purpose of this paper is precisely to address this issue.

In this present manuscript, our aim is to obtain the rational extension of 
all the non-central 
conventional potentials discussed in Ref. \cite{kb}. Further, one of the major
development after the work of \cite{kb} has been the discovery of $PT$-invariant
complex potentials with real energy eigenvalues. 
The concept of $PT$ (combined parity ($P$) and time reversal ($T$))
symmetric quantum mechanics \cite{bender} also plays a crucial role in the understanding of the complex quantum mechanical systems. Bender et.al., 
\cite{bender} have showed that the eigenspectrum of such non-hermitian 
$PT$ symmetric complex systems are real provided the PT-symmetry is not
spontaneously broken. The second purpose of this paper is to consider 
$PT$-symmetric, noncentral, exactly solvable, potentials  and obtain their
rational extensions. It turns out that the bound state eigenfunctions of some of these 
potentials are not in the exact form of EOPs rather they are written 
in the form of some new types of polynomials which can be further written in 
terms of the corresponding classical orthogonal polynomials.

The paper is organized as follows:

 In section $2$, we start from the non-central potentials with seven 
parameters (six continuous and one discrete)
in spherical polar co-ordinates and explain how one can obtain its 
rationally extended solution in terms of ten parameters (six continuous and
four discrete) solutions. Various forms of $r$, $\theta$ and 
$\phi$ dependent potential terms are also mentioned. For an illustration, 
we discuss in detail in Sec. $2.1$ and $2.2$ respectively two examples (one real 
and one complex and $PT$ symmetric) of  ten parameters 
 rationally extended non-central potentials. In particuar, we show that 
the corresponding eigenfunctions are product of the exceptional Jacobi,  
 exceptional  Laguerre and/or some type of new orthogonal polynomials. 
A list of possible forms of $\theta$ and $\phi$ dependent terms with the 
corresponding eigenfunctions as well as the 
other parametric relations are given in Tables $I$ and $II$ respectively. 
Some examples of RE non-central potentials in two dimensional polar 
co-ordinates are also mentioned in brief in Sec. $3$. In particular, we start
from the five parameter (four continuous and one discrete) families of
non-central potentials in polar coordinates and obtain the corresponding
rationally extended non-central potentials with seven parameters (four
continuos and three discrete).  Finally, we summarize 
our results in section $4$. 
 
\section{Non central potential in $3$-dimensional spherical polar co-ordinates}
In spherical polar coordinates $(r,\theta,\phi)$, consider  a non central 
potential \cite{lif_76} of the form  
as 
\be\label{non_form}
V(r,\theta,\phi)=\tilde{U}(r)+\frac{V(\theta)}{r^2}+\frac{U(\phi)}{r^2\sin^2 (\theta)}.
\ee 
The Schr\"odinger equation corresponding to this potential i.e.,
\ba\label{sch_non}
\bigg[-\bigg(\frac{\partial^2\Psi}{\partial r^2}+\frac{2}{r}\frac{\partial\Psi}{\partial r}\bigg)- \frac{1}{r^2}\bigg(\frac{\partial^2\Psi}{\partial \theta^2}+\cot\theta \frac{\partial\Psi}{\partial\theta}\bigg)-\frac{1}{r^2\sin^2 \theta}\frac{\partial^2 \Psi}{\partial \phi^2}\bigg]=(E-V(r,\theta,\phi))\Psi,
\ea
has been solved exactly \cite{kb} by using the fact that the eigenfunction
can be written in the product form 
\be\label{wf_non}
\Psi(r,\theta,\phi)=\frac{R(r)}{r}\frac{\Theta(\theta)}{(\sin\theta)^{\frac{1}{2}}}\Phi(\phi).
\ee 
Using Eq. (\ref{wf_non}) in Eq. (\ref{sch_non}), we obtain three exactly 
solvable uncoupled equations given by 
\be\label{un_cop_1}
-\frac{\partial^2 \Phi(\phi)}{\partial \phi^2}+U(\phi) \Phi(\phi)=m^2\Phi(\phi),
\ee     
\be\label{un_cop_2}
-\frac{\partial^2 \Theta(\theta)}{\partial \theta^2}+\bigg[V(\theta)+\bigg(m^2-\frac{1}{4}\bigg)\cosec^2\theta\bigg] \Theta(\theta)=\ell^2\Theta(\theta)
\ee     
and 
\be\label{un_cop_3}
-\frac{\partial^2 R(r)}{\partial r^2}+\bigg[\tilde{U}(r)+\frac{(\ell^2-1/4)}{r^2}\bigg]R(r)=ER(r).
\ee     
By considering different forms of a seven parameter family of potentials, 
 the above three Eqs. (\ref{un_cop_1}) - (\ref{un_cop_3}) have been  
solved exactly \cite{kb} and in this way one has constructed several 
non-central potentials 
$V(r,\theta,\phi)$ by considering different forms of $\tilde{U}(r), U(\theta)$ 
and $U(\phi)$ with the corresponding 
eigenvalues being $E, \ell^2$ and $m^2$ respectively. The eigenfunctions corresponding to these three equations (\ref{un_cop_1}) - (\ref{un_cop_3})
 are either in terms of classical 
Laguerre or Jacobi orthogonal polynomials. The complete eigenfunctions are 
obtained by using Eq. (\ref{wf_non}). The  
forms of the potential $V(r,\theta,\phi)$ with their solutions can be found 
in detail in Ref. \cite {kb}. As an illustration, one of the seven parameter
family of potential considered in \cite{kb} is given by 
\ba
V(r,\theta,\phi)&=&\frac{\omega^2 r^2}{4}+\frac{\delta}{r^2}
+\frac{C}{r^2\sin^2\theta}+\frac{D}{r^2\cos^2\theta} \nonumber\\
&+&\frac{G}{r^2\sin^2\theta\sin^2 p\phi}
 +\frac{F}{r^2\sin^2\theta \cos^2p \phi}\,, 
\ea

In this work, If we change the potential $V(r,\theta,\phi) \Rightarrow V_{m_1,m_2,m_3}(r,\theta,\phi)$
by redefining the extended form $\tilde{U}(r)\Rightarrow \tilde{U}_{m_1,ext}(r)$, $V(\theta)\Rightarrow V^{(h)}_{m_2,ext}(\theta)$,  $U(\phi)\Rightarrow U^{(h)}_{m_3,ext}(\phi)$
i.e.,
\ba\label{v_rat}
V(r,\theta,\phi) &\Rightarrow &V_{m_1,m_2,m_3}(r,\theta,\phi)\nonumber\\
&=&\tilde{U}_{m_1,ext}(r)+\frac{1}{r^2}V^{(h)}_{m_2,ext}(\theta)+\frac{1}{r^2\sin^2\theta}U^{(h)}_{m_3,ext}(\phi),
\ea 
and the eigenfunction 
\ba\label{wf_non_rat}
\Psi(r,\theta,\phi)&\Rightarrow&\Psi_{m_1,m_2,m_3}(r,\theta,\phi)\nonumber\\
&=& \frac{R_{m_1}(r)}{r}\frac{\Theta^{(h)}_{m_2}(\theta)}{(\sin\theta)^{\frac{1}{2}}}\Phi^{(h)}_{m_3}(\phi),
\ea
then Eq. (\ref{sch_non}) becomes
\ba\label{sch_rat}
&&\bigg[-\frac{1}{R_{m_1}}\frac{\partial^2R_{m_1}}{\partial r^2}+\tilde{U}_{m_1,ext}(r)-\frac{1}{4r^2}\bigg]+\frac{1}{r^2}\bigg[-\frac{1}{\Theta^{(h)}_{m_2}}\frac{\partial^2\Theta^{(h)}_{m_2}}{\partial \theta^2}+V^{(h)}_{m_2,ext}(\theta)-\frac{1}{4}\cosec^2\theta\bigg]\nonumber\\
&&+\frac{1}{r^2\sin^2\theta}\bigg[-\frac{1}{\Phi^{(h)}_{m_3}}\frac{\partial^2\Phi^{(h)}_{m_3}}{\partial \phi^2}+U^{(h)}_{m_3,ext}(\phi)\bigg]=E.\nonumber\\
\ea
Here $h=I, II$ correspond to real potential while $h= (PT)_1, (PT)_2$ correspond to the complex and $PT$ 
symmetric potential in $\theta$ and/or $\phi$.

Similar to Eqs. (\ref{un_cop_1})-(\ref{un_cop_3}), the above Eq. (\ref{sch_rat}) can also be easily uncoupled into three new exactly 
solvable equations given by
\be\label{ucext_1}
-\frac{\partial^2 \Phi^{(h)}_{m_3}(\phi)}{\partial \phi^2}+U^{(h)}_{m_3,ext}(\phi)\Phi^{(h)}_{m_3}(\phi)=m^2\Phi^{(h)}_{m_3}(\phi),
\ee     
\be\label{ucext_2}
-\frac{\partial^2 \Theta^{(h)}_{m_2}(\theta)}{\partial \theta^2}+\bigg[V^{(h)}_{m_2,ext}(\theta))+\bigg(m^2-\frac{1}{4}\bigg)\cosec^2\theta\bigg] \Theta^{(h)}_{m_2}(\theta)=\ell^2\Theta^{(h)}_{m_2}(\theta)
\ee     
and 
\ba\label{ucext_3}
-\frac{\partial^2 R_{m_1}(r)}{\partial r^2}+\bigg[\tilde{U}_{m_1,ext}(r)+\frac{(\ell^2-1/4)}{r^2}\bigg]R_{m_1}(r)=ER_{m_1}(r). 
\ea     
By knowing the solutions of the above Eqs. (\ref{ucext_1})-(\ref{ucext_3}), a complete solution of 
the extended non-central potential $V_{m_1,m_2,m_3}(r,\theta,\phi)$ can be obtained by using Eq. (\ref{wf_non_rat}) with the 
energy eigenvalues $E$. 

 We show that there is one choice of $\tilde{U}_{m_1,ext}(r)$, four choices of $V^{(h)}_{m_2,ext}(\theta)$ (two real and two $PT$ symmetric) and three choices of 
 $U^{(h)}_{m_3,ext}(\phi)$ (two real and one $PT$ symmetric) for which one can
obtain exact solutions of the non-central potential. These choices are:

{\bf {$(a)$ Form of $\tilde{U}_{m_1,ext}(r)$:}}
\be\label{ext_r}
\tilde{U}_{m_1,ext}(r)=\frac{\omega^2 r^2}{4}+\frac{\delta}{r^2}+\tilde{U}_{m_1,rat}(r),
\ee
where
\ba\label{rat1}
\tilde{U}_{m_1,rat}(r)&=&-2m_1\omega-\omega^2r^2\frac{L^{(\tilde{\delta}+1)}_{m_1-2}(-\omega r^2/2)}{L^{(\tilde{\delta}-1)}_{m_1}(-\omega r^2/2)}\nonumber\\
&+&\omega(\omega r^2+2(\tilde{\delta}-1))\frac{L^{(\tilde{\delta})}_{m_1-1}(-\omega r^2/2)}{L^{(\tilde{\delta}-1)}_{m_1}(-\omega r^2/2)}\nonumber\\
&+& 2\omega^2 r^2\bigg(\frac{L^{(\tilde{\delta})}_{m_1-1}(-\omega r^2/2)}{L^{(\tilde{\delta}-1)}_{m_1}(-\omega r^2/2)}\bigg)^2.\nonumber
\ea
Here $L^{(\tilde{\delta})}_{m_1}(-\omega r^2/2)$ is a classical Laguerre polynomial.

{\bf {$(b)$ Forms of $V^{(h)}_{m_2,ext}(\theta)$:}}

$(i)$ For $h=I$
\be\label{ext_theta_1}
V^{(I)}_{m_2,ext}(\theta)=\frac{C}{\sin^2 \theta}+\frac{D}{\cos^2 \theta}+V^{(I)}_{m_2,rat}(\theta),
\ee
where the rational part $V^{(I)}_{m_2,rat}(\theta)$ is given by
\ba\label{rat2}
 V^{(I)}_{m_2,rat}(\theta)&=&4\bigg[-2m_2(\alpha -\beta -m_2+1)-(\alpha -\beta -m_2+1)\big(\alpha +\beta +(\alpha -\beta +1)\cos(2\theta)\big)\nonumber\\
&\times & \frac{P_{m_2-1}^{(-\alpha ,\beta) }(\cos (2\theta))}{P_{m_2}^{(-\alpha -1,\beta -1)}(\cos (2\theta))}+\frac{(\alpha -\beta -m_2+1)^{2}\sin^2(2\theta)}{2}
\bigg(\frac{P_{m_2-1}^{(-\alpha ,\beta) }(\cos(2\theta))}{P_{m_2}^{(-\alpha -1,\beta -1)}(\cos(2\theta))}\bigg)^{2}\bigg].\nonumber 
\ea  

$(ii)$ For $h=II$
\be\label{ext_theta_2}
V^{(II)}_{m_2,ext}(\theta)=\frac{C}{\sin^2 \theta}+\frac{D}{\sin\theta \tan \theta}+V^{(II)}_{m_2,rat}(\theta),
\ee
with the rational part
\ba\label{rat_theta2}
V_{m_2,rat}^{(II)}(\theta ) &=& \bigg[-2 m_{2}(\alpha-\beta - m_{2} + 1 ) - 
(\alpha-\beta - m_{2} +1)(\alpha + \beta + (\alpha-\beta + 1 )\cos \theta)\nonumber \\
&\times &\left(\frac{P_{m_{2}-1}^{(-\alpha , \beta)} (\cos \theta)} 
{P_{m_{2}}^{(-\alpha-1,\beta -1)}(\cos \theta ) }\right) 
 + \frac{(\alpha-\beta -m_{2}+1 )^{2}\sin^{2}\theta}{2}\left(\frac{P_{m_{2}-1}^{(-\alpha,\beta )} (\cos \theta)} 
{P_{m_{2}}^{(-\alpha-1, \beta -1)}(\cos \theta) }\right)^{2} \bigg].\nonumber
\ea  

$(iii)$ For $h=(PT)_1$

\be\label{ext_theta_pt}
V^{(PT)_1}_{m_2,ext}(\theta)=\frac{C}{\sin^2 \theta}+\frac{iD}{\tan \theta}+V^{(PT)_1}_{m_2,rat}(\theta),
\ee
is a complex and $PT$ symmetric form of $V^{(h)}_{m_2,ext}(\theta)$ with the corresponding complex and $PT$ symmetric rational term
\ba
V^{(PT)_1}_{m_2,rat}(\theta) &=& -2\cosec^2\theta\bigg[ 2i\cot\theta \frac{\dot{q}^{(A,B)}_{m_2}(z)}{q^{(A,B)}_{m_2}(z)}-\cosec^2\theta\nonumber\\
&\times & \Bigg( \frac{\ddot{q}^{(A,B)}_{m_2}(z)}{q^{(A,B)}_{m_2}(z)}-\bigg( \frac{\dot{q}^{(A,B)}_{m_2}(z)}{q^{(A,B)}_{m_2}(z)}\bigg )^2 \Bigg )-m_2\bigg ].\nonumber
\ea 

$(iv)$ For $h=(PT)_2$

The potential given in case (ii) can also be made complex and $PT$ symmetric by multiplying the potential parameter $D$ by imaginary number $i$ and get
\be\label{ext_theta_pt2}
V^{(PT)_2}_{m_2,ext}(\theta)=\frac{C}{\sin^2 \theta}+\frac{iD}{\sin\theta \tan \theta}+V^{(PT)_2}_{m_2,rat}(\theta),
\ee
with the complex rational part
\ba\label{rat_theta_pt2}
V_{m_2,rat}^{(PT)_2}(\theta ) &=& \bigg[-2 m_{2}(\alpha-\beta - m_{2} + 1 ) - 
(\alpha-\beta - m_{2} +1)(\alpha + \beta + (\alpha-\beta + 1 )\cos \theta)\nonumber \\
&\times &\left(\frac{P_{m_{2}-1}^{(-\alpha , \beta)} (\cos \theta)} 
{P_{m_{2}}^{(-\alpha-1,\beta -1)}(\cos \theta ) }\right) 
 + \frac{(\alpha-\beta -m_{2}+1 )^{2}\sin^{2}\theta}{2}\left(\frac{P_{m_{2}-1}^{(-\alpha,\beta )} (\cos \theta)} 
{P_{m_{2}}^{(-\alpha-1, \beta -1)}(\cos \theta) }\right)^{2} \bigg].\nonumber
\ea  
Here the potential parameters $\alpha$ and $\beta$ are complex.

{\bf {$(c)$ Forms of $V^{(h)}_{m_3,ext}(\phi)$:}}

$(i)$ For $h=I$
\be\label{ext_phi_1}
V^{(I)}_{m_3,ext}(\phi)=\frac{G}{\sin^2 (p\phi )}+\frac{F}{\cos^2 (p\phi)}+V^{(I)}_{m_3,rat}(\phi),
\ee
where
\ba\label{rat3}
 U^{(I)}_{m_3,rat}(\phi)&=&4p^2\bigg[-2m_3(\tilde{\alpha}-\tilde{\beta} -m_3+1)-(\tilde{\alpha} -\tilde{\beta} -m_3+1)\big(\tilde{\alpha} +\tilde{\beta} +(\tilde{\alpha} -\tilde{\beta} +1)\cos(2p\phi)\big)\nonumber\\
&\times & \frac{P_{m_2-1}^{(-\tilde{\alpha} ,\tilde{\beta}) }(\cos (2p\phi))}{P_{m_2}^{(-\tilde{\alpha} -1,\tilde{\beta} -1)}(\cos (2p\phi))}+\frac{(\tilde{\alpha} -\tilde{\beta} -m_2+1)^{2}}{2}\sin^2(2p\phi)
\bigg(\frac{P_{m_2-1}^{(-\tilde{\alpha} ,\tilde{\beta}) }(\cos(2p\phi))}{P_{m_2}^{(-\tilde{\alpha} -1,\tilde{\beta} -1)}(\cos(2p\phi))}\bigg)^{2}\bigg].\nonumber 
\ea 

Note that here $p$ is any positive integer.

$(ii)$ For $h=II$
\be\label{ext_phi_2}
V^{(II)}_{m_3,ext}(\phi)=\frac{G}{\sin^2 (p\phi)}+\frac{F}{\sin(p\phi) \tan(p\phi)}+V^{(II)}_{m_3,rat}(\phi),
\ee
where
\ba
U_{m_3,rat}^{(II)}(\phi ) &= &  p^{2}\bigg[-2 m_{3}(\tilde{\alpha }- \tilde{\beta } - m_{3} + 1 ) - 
(\tilde{\alpha }-\tilde{\beta } - m_{3} +1)(\tilde{\alpha } + \tilde{\beta }+ (
\tilde{\alpha }-\tilde{\beta } + 1 )\cos ( p\phi ) )\nonumber \\
 &\times &\left(\frac{P_{m_{3}-1}^{(-\tilde{\alpha }, \tilde{\beta })} (\cos (p\phi ))} 
{P_{m_{3}}^{(-\tilde{\alpha }-1, \tilde{\beta } -1)}(\cos ( p\phi )) }\right) 
 + \frac{(\tilde{\alpha }-\tilde{\beta } -m_{3}+1 )^{2}\sin^{2}(p\phi )}{2}\nonumber\\
&\times &\left(\frac{P_{m_{3}-1}^{(-\tilde{\alpha }, \tilde{\beta })} (\cos (p\phi ))} 
{P_{m_{3}}^{(-\tilde{\alpha }-1, \tilde{\beta } -1)}(\cos ( p\phi )) }\right)^{2} \bigg].\nonumber
\ea

Here $p$ is any odd positive integer.

$(iii)$ For $h=(PT)_1$

The above case (ii) of the $\phi$ dependent terms can be made complex and $PT$ symmetric by replacing the potential parameter $F\rightarrow iF$ and 
we get  
\be\label{ext_phi_3}
V^{(PT)}_{m_3,ext}(\phi)=\frac{G}{\sin^2 (p\phi)}+\frac{iF}{\sin(p\phi) \tan(p\phi)}+V^{(PT)_1}_{m_3,rat}(\phi),
\ee
where the complex rational term with the complex parameters $\tilde{\alpha}$ and $\tilde{\beta}$  given by 
\ba
U_{m_3,rat}^{(PT)_1}(\phi ) &= &  p^{2}\bigg[-2 m_{3}(\tilde{\alpha }- \tilde{\beta } - m_{3} + 1 ) - 
(\tilde{\alpha }-\tilde{\beta } - m_{3} +1)(\tilde{\alpha } + \tilde{\beta }+ (
\tilde{\alpha }-\tilde{\beta } + 1 )\cos ( p\phi ) )\nonumber \\
 &\times &\left(\frac{P_{m_{3}-1}^{(-\tilde{\alpha }, \tilde{\beta })} (\cos (p\phi ))} 
{P_{m_{3}}^{(-\tilde{\alpha }-1, \tilde{\beta } -1)}(\cos ( p\phi )) }\right) 
 + \frac{(\tilde{\alpha }-\tilde{\beta } -m_{3}+1 )^{2}\sin^{2}(p\phi )}{2}\nonumber\\
&\times &\left(\frac{P_{m_{3}-1}^{(-\tilde{\alpha }, \tilde{\beta })} (\cos (p\phi ))} 
{P_{m_{3}}^{(-\tilde{\alpha }-1, \tilde{\beta } -1)}(\cos ( p\phi )) }\right)^{2} \bigg].\nonumber
\ea
Note that here $p$ is any odd positive integer.

Here $P^{(\tilde{\alpha},\tilde{\beta})}_{m_2}(z)$  and $P^{(\alpha,\beta)}_{m_3}(z)$ are classical Jacobi polynomials. 

 By taking various combinations of these  allowed choices, we then have twelve 
different, rational, exactly solvable non-central
potentials in three dimensions, each with ten parameters. These choices of 
potentials, the
corresponding eigenvalues and eigenfunctions are given in Tables $I$ and $II$.
In particular, in Table $I$ we give the four forms of possible $V^{(h)}_{m_2,ext}(\theta)$ 
(two real and two complex but $PT$-invariant) with their corresponding eigenvalues
and eigenfunctions. In Table $II$, we similarly give two real
and one complex and $PT$ symmetric forms of $U^{(h)}_{m_3,ext}(\phi)$ and the 
corresponding eigenvalues and eigenfunctions.

As an illustration, we discuss two examples in detail, one real and one 
$PT$-invariant complex case in Secs. $2.1$ and $2.2$ respectively.

\subsection{Example of Rationally Extended (RE) non-central real potential}
In this section, we discuss an example of the ten parameters (six continuous 
and four discrete) RE non-central real potential and its bound state solutions 
explicitly. We consider the potential of the form 

\ba\label{non_ten}
V_{m_1,m_2,m_3}(r,\theta,\phi)&=&\frac{\omega^2 r^2}{4}+\frac{\delta}{r^2}+\tilde{U}_{m_1,rat}(r)+\frac{C}{r^2\sin^2\theta}+\frac{D}{r^2\cos^2\theta}+\frac{1}{r^2}V^{(I)}_{m_2,rat}(\theta)\nonumber\\
&+&\frac{G}{r^2\sin^2\theta\sin^2 p\phi}
 +\frac{F}{r^2\sin^2\theta \cos^2p \phi}+\frac{1}{r^2\sin^2\theta}U^{(I)}_{m_3,rat}(\phi),  
\ea
 where the six parameters $\omega, \delta, C, D, F$ and $ G $  are continuous 
parameters while the rest 
four i.e., $p, m_1, m_2$ and $m_3$ are discrete parameters. In particular, each
of them can take any inegral value. The rational terms $\tilde{U}_{m_1,rat}(r), V^{(I)}_{m_2,rat}(\theta)$ and $U^{(I)}_{m_3,rat}(\phi)$
are given by Eqs. (\ref{ext_r}), (\ref{ext_theta_1}) and (\ref{ext_phi_1}) 
respectively. It is easy to show that the eigenvalues of this extended 
non-central potential are the same as that of the conventional 
case given by Eq. (\ref{sch_non}) but the eigenfunctions are  different 
which are obtained in terms of EOPs. The complete eigenfunction is given by Eq. (\ref{wf_non_rat}) 
which is ultimately a product of these EOPs. 
To solve the above extended non-central potential, first we consider 
 a simple case of $m_1=m_2=m_3=1$ and then we generalize it to any arbitrary 
positive integer values of $m_1,m_2$ and $ m_3$.

{\bf Case (i): For $m_1=m_2=m_3=1$}

In this case, the ten parameters RE non-central potential is reduced to 
a seven parameters RE non-central potential 
 \ba\label{pot_extd_1}
 V_{1,1,1}(r,\theta,\phi)&=&\frac{\omega^2 r^2}{4}+\frac{\delta}{r^2}+\tilde{U}_{1,rat}(r)+\frac{C}{r^2\sin^2\theta}+\frac{D}{r^2\cos^2\theta}+\frac{1}{r^2}V^{(I)}_{1,rat}(\theta)\nonumber\\
&+&\frac{G}{r^2\sin^2\theta\sin^2 p\phi}
 +\frac{F}{r^2\sin^2\theta \cos^2p \phi}+\frac{1}{r^2\sin^2\theta}U^{(I)}_{1,rat}(\phi),
  \ea
  here $p$ is any positive integer. To get the exact solution of the above 
Eq. (\ref{pot_extd_1}), we define the rational terms 
$\tilde{U}_{1,rat}(r), V^{(I)}_{1,rat}(\theta)$ and $U^{(I)}_{1,rat}(\phi)$ as (by putting $m_1=m_2=m_3$ in the rational parts of Eqs. 
(\ref{ext_r}), (\ref{ext_theta_1}) and (\ref{ext_phi_1}))
\ba\label{prad_1}
\tilde{U}_{1,rat}(r)=\frac{4\omega}{(\omega r^2+2\tilde{\delta})}-\frac{16\omega \tilde{\delta}}{(\omega r^2+2\tilde{\delta})^2},
\ea
\ba\label{ptheta_1}
V^{(I)}_{1,rat}(\theta)=\frac{8(\alpha+\beta)}{\big ((\alpha+\beta )-(\beta-\alpha )\cos ( 2\theta ) \big)}-\frac{8\big((\alpha+\beta)^2-(\beta-\alpha)^2\big)}{\big((\alpha+\beta )-(\beta-\alpha )\cos ( 2\theta ) \big)^2},
\ea
and 
\ba\label{pphi_1}
U^{(I)}_{1,rat}(\phi)=4p^2\bigg[\frac{2(\tilde{\alpha}+\tilde{\beta})}{\big ((\tilde{\alpha}+\tilde{\beta} )-(\tilde{\beta}-\tilde{\alpha} )\cos ( 2p\phi ) \big)}-\frac{2\big((\tilde{\alpha}+\tilde{\beta})^2-(\tilde{\beta}-\tilde{\alpha})^2\big)}{\big((\tilde{\alpha}+\tilde{\beta} )-(\tilde{\beta}-\tilde{\alpha} )\cos ( 2p\phi ) \big)^2}\bigg].
\ea
On comparing Eq. (\ref{pot_extd_1}) with Eq. (\ref{v_rat} ) (for $m_1=m_2=m_3=1$ and $h=I$), we get the rationally extended 
trigonometric P\"oschl-Teller potential \cite{que,os}  
\ba\label{extd_phi_1}
U^{(I)}_{1,ext}(\phi)= U^{(I)}_{con}(\phi)+ U^{(I)}_{1,rat}(\phi),
\ea
where 
\be\label{con_rpt} 
U^{I}_{con}(\phi) = G \cosec^2 (p\phi) + F\sec^2(p\phi)
\ee
is the corresponding conventional potential. The unnormalized $\phi$ dependent wave function of Eq. (\ref{ucext_1}) (for $m_3=1$) 
with the extended potential (\ref{extd_phi_1}) in terms of $X_1$ exceptional orthogonal polynomials $\hat{P}_{n_3+1}^{(\tilde{\alpha} ,\tilde{\beta}) }(z)$ is well known and given by \cite {que,os}
\be\label{ext_phi_sol_1}
\Phi^{I}_{1,n_3}(\phi)\propto \frac{(1-z )^{\frac{\tilde{\alpha} }{2} + \frac{1}{4}}(1 +z )^{\frac{\tilde{\beta}  }{2} + \frac{1}{4}}}
{\big ((\tilde{\alpha}+\tilde{\beta} )-(\tilde{\beta}-\tilde{\alpha} )\cos ( 2p\phi ) \big)}\hat{P}_{n_3+1}^{(\tilde{\alpha} ,\tilde{\beta}) }(z);\qquad 0\leq p\phi\leq \pi/2,
\ee
where $n_3=0,1,2,3...$, \quad $z=\cos(2p\phi)$ and the positive constant parameters
\ba\label{al_telda}
\tilde{\alpha} =\frac{1}{2}\sqrt {1+\frac{4G}{p^2}};\quad \tilde{\beta} &=& \frac{1}{2}\sqrt {1+\frac{4F}{p^2}}.
\ea 
 The eigenvalue spectrum of this extended potential is same (i.e. isospectral) 
as that of the conventional 
potential $U_{con}(\phi)$ which is given by  
\be\label{ms}
m^2=p^2(2n_3+\tilde{\alpha}+\tilde{\beta}+1)^2.
\ee 
Again from Eqs. (\ref{pot_extd_1}) and (\ref{v_rat}), the rationally extended $\theta$ dependent potential is given by
 \ba\label{extd_theta_1}
V^{(I)}_{1,ext}(\theta)= V^{(I)}_{con}(\theta)+ V^{(I)}_{1,rat}(\theta),
\ea
where the conventional potential
\be\label{con_tpt} 
V^{(I)}_{con}(\theta) = C \cosec^2(\theta) + D\sec^2(\theta).
\ee
The Schr\"odinger equation  (\ref{ucext_2}) (for $m_2=1$ and $h=I$) becomes
\ba\label{sch_theta_1}
-\frac{\partial^2 \Theta^{(I)}_{1}(\theta)}{\partial \theta^2}+\bigg[\bigg(C+m^2-\frac{1}{4}\bigg)\cosec^2\theta+D\sec^2\theta+ V^{(I)}_{1,rat}(\theta)\bigg] \Theta^{(I)}_{1}(\theta)=\ell^2\Theta^{(I)}_{1}(\theta).
\ea
Using the rational term $V^{(I)}_{1,rat}(\theta)$ from Eq. (\ref{ptheta_1}), the wave function and the eigenspectrum of the above  Eq. (\ref{sch_theta_1}) are thus given by 
\be\label{ext_theta_sol_1}
\Theta^{(I)}_{1,n_2}(\theta)\propto \frac{(1-z )^{\frac{\alpha}{2} + \frac{1}{4}}(1 +z )^{\frac{\beta}{2} + \frac{1}{4}}}
{\big ((\alpha+\beta )-(\beta-\alpha )\cos ( 2\theta ) \big)}\hat{P}_{n_2+1}^{(\alpha ,\beta) }(z);\qquad 0\leq \theta \leq \pi/2,
\ee
and 
\be\label{ls}
\ell^2=(2n_2+\alpha+\beta+1)^2; \qquad n_2=0,1,2,..., 
\ee  
where $z=\cos(2\theta)$ and the parameters
\ba\label{alpha}
\alpha &=& \sqrt {C+m^2},\nonumber\\
\beta &=& \frac{1}{2}\sqrt {1+4D}.
\ea
Note that the eigenvalue spectrum is unchanged while the eigenfunctions are
different from those of the nonrational case.

Similar to the above cases, from Eqs. (\ref{pot_extd_1}) and (\ref{v_rat}), the radial component of the extended potential is given by
 \ba\label{extd_rad_1}
\tilde{U}_{1,ext}(r)=\tilde{U}_{con}(r) + \tilde{U}_{1,rat}(r),
\ea
where the conventional radial oscillator potential
\be\label{con_rad}
\tilde{U}_{con}(r) = \frac{\omega^2 r^2}{4}+\frac{\delta}{r^2}.
\ee
From Eq. (\ref{ucext_3}) (for $m_1=1$), finally, we get the exactly solvable Schr\"odinger equation 
\be
-\frac{\partial^2 R_{1}(r)}{\partial r^2}+\bigg[\frac{\omega^2 r^2}{4}+\frac{(\delta+\ell^2-1/4)}{r^2}+\tilde{U}_{1,rat}(r)\bigg]R_{1}(r)=ER_{1}(r),
\ee
with the solution in term of $X_1$ Laguerre EOPs $\hat{L}_{n_1+1}^{(\tilde{\delta}) }\big(\frac{\omega r^2}{2}\big)$ given by \cite{que,os}
\be\label{ext_rad_sol_1}
R_{1,n_1}(r)\propto \frac{r^{(\tilde{\delta}+1/2)}\exp{\big(-\frac{\omega r^2}{4}\big)}}
{(\omega r^2+2\tilde{\delta})}\hat{L}_{n_1+1}^{(\tilde{\delta}) }\big(\frac{\omega r^2}{2}\big);\qquad 0< r < \infty.
\ee
The energy eigenvalue $E$ which depends on $n_1, n_2$ and $n_3$ is given by
\be\label{en123}
E_{n_1,n_2,n_3}=\omega(2n_1+1+\tilde{\delta}), 
\ee  
where $\tilde{\delta}=\sqrt{\delta+\ell^2}$ and 
\be\label{en_1}
\ell^2=\bigg[(2n_2+1)+\sqrt{D+1/4}+\bigg \{ C+\bigg(\sqrt{F+p^2/4}+\sqrt{G+p^2/4}+p(2n_3+1)\bigg)^2\bigg \}^{1/2}\bigg]^2.
\ee
Again note that the energy eigenvalues are unchanged while the corresponding
eigenfunctions are different from the nonrational case.

 {\bf Case (ii): For any positive integer values of $m_1, m_2$ and $m_3$}

In this case, we consider a more general form of the ten parameters potential given in Eq. (\ref{non_ten}). Similar to the $X_1$ case, on comparing Eqs. (\ref{non_ten}) and (\ref{v_rat}), we obtain the 
rationally extended trigonometric P\"oschl-Teller equation
\ba\label{extd_phi}
U^{(I)}_{m_3,ext}(\phi)= U^{(I)}_{con}(\phi)+ U^{(I)}_{m_3,rat}(\phi),
\ea
where $U^{(I)}_{con}(\phi)$ and $U^{(I)}_{m_3,rat}(\phi)$ are given by Eqs. (\ref{con_rpt}) and (\ref{rat3}).
The unnormalized wavefunction of Eq. (\ref{ucext_1}) with the extended potential (\ref{extd_phi}) in terms of 
the $X_{m_3}$ exceptional Jacobi polynomial $\hat{P}_{n_3+m_3}^{(\tilde{\alpha} ,\tilde{\beta}) }(z)$ is given by \cite {os}
\be\label{ext_phi_sol}
\Phi^{(I)}_{m_3,n_3}(\phi)\propto \frac{(1-z )^{\frac{\tilde{\alpha} }{2} + \frac{1}{4}}(1 +z )^{\frac{\tilde{\beta}  }{2} + \frac{1}{4}}}
{P_{m_3}^{(-\tilde{\alpha} -1,\tilde{\beta} -1)}(z)}\hat{P}_{n_3+m_3}^{(\tilde{\alpha} ,\tilde{\beta}) }(z);\qquad 0\leq p\phi\leq \pi/2,
\ee
where $n_3=0,1,2,....;$ \quad $m_3=1,2,3...$; \quad $z=\cos(2p\phi)$ and the parameters
$\tilde{\alpha}$ and $\tilde{\beta}$ will be same as obtained in Eq. (\ref{al_telda}). 
 The eigenvalue of this extended potential is same (i.e. isospectral) as that of the conventional 
potential $U^{(I)}_{con}(\phi)$ given by Eq. (\ref{ms}).  
 
Similar to the above case from Eqs. (\ref{non_ten}) and (\ref{v_rat}), the $\theta$ and $r$ dependent 
extended potentials which depend on parameters 
$m_2$ and $m_3$ respectively are given as 
 \ba\label{extd_theta}
V^{(I)}_{m_2,ext}(\theta)= V^{(I)}_{con}(\theta)+ V^{(I)}_{m_2,rat}(\theta),
\ea
and 
\ba\label{extd_rad}
\tilde{U}_{m_1,ext}(r)=\tilde{U}_{con}(r) + \tilde{U}_{m_1,rat}(r).
\ea
The corresponding conventional potential terms  $V^{(I)}_{con}(\theta)$ and $\tilde{U}_{con}(r)$ and the rational terms $V^{(I)}_{m_2,rat}(\theta)$ and $\tilde{U}_{m_1,rat}(r)$ are given in Eqs. (\ref{extd_theta_1}), (\ref{extd_rad_1}) and (\ref{rat2}), (\ref{rat1}).
Following the same procedure as in the $X_1$ case, the solutions of the the Eqs. (\ref{ucext_2}) and (\ref{ucext_3}) with the corresponding 
potentials $V^{(I)}_{m_2,ext}(\theta)$ and $\tilde{U}_{m_1,ext}(r)$ can be obtained in a simple way. Thus the bound state wavefunctions 
corresponding to these potentials are given by     
\be\label{ext_theta_sol}
\Theta^{(I)}_{m_2,n_2}(\theta)\propto \frac{(1-z )^{\frac{\alpha}{2} + \frac{1}{4}}(1 +z )^{\frac{\beta}{2} + \frac{1}{4}}}
{P_{m_2}^{(-\alpha -1,\beta -1)}(z)}\hat{P}_{n_2+m_2}^{(\alpha ,\beta) }(z);\qquad 0\leq \theta \leq \pi/2,
\ee
and 
\be\label{ext_rad_sol}
R_{m_1,n_1}(r)\propto \frac{r^{(\sqrt{\delta+\ell^2}+1/2)}\exp{\big(-\frac{\omega r^2}{4}\big)}}
{L_{m_1}^{(\sqrt{\delta+\ell^2}-1)}\big(-\frac{\omega r^2}{2}\big)}\hat{L}_{n_1+m_1}^{(\sqrt{\delta+\ell^2}) }\big(\frac{\omega r^2}{2}\big);\qquad 0< r < \infty,
\ee
where $\hat{P}_{n_2+m_2}^{(\alpha ,\beta) }(z)$ and $\hat{L}_{n_1+m_1}^{(\sqrt{\delta+\ell^2}) }\big(\frac{\omega r^2}{2}\big)$ are 
$X_{m_2}$ exceptional Jacobi and $X_{m_1}$ exceptional Laguerre polynomials respectively. The energy eigenspectrum and the other parametric 
relations will be same as that of the $X_1$ case.

 \subsection{ Example of Rationally Extended $PT$ symmetric complex non-central potential}

Similar to the above example of real case, here we consider an example of RE 
non-central potential which is complex but symmetric under the combined 
operation of the parity ($P$) ($P: r\rightarrow r, \theta\rightarrow \pi-\theta, \phi\rightarrow \phi+\pi$) and the 
time reversal ($T$) ($T: t \rightarrow -t, i\rightarrow -i$) operators and given by  


\ba\label{pot_extd_pt}
 V^{(PT)}_{m_1,m_2,m_3}(r,\theta,\phi)&=&\frac{\omega^2 r^2}{4}+\frac{\delta}{r^2}+\tilde{U}_{m_1,rat}(r)+\frac{C}{r^2\sin^2\theta}+\frac{iD}{r^2\tan\theta}+\frac{1}{r^2}V^{(PT)}_{m_2,rat}(\theta)\nonumber\\
&+&\frac{G}{r^2\sin^2\theta\sin^2 p\phi}
 +\frac{F}{r^2\sin^2\theta \cos^2p \phi}+\frac{1}{r^2\sin^2\theta}U^{(I)}_{m_3,rat}(\phi),
  \ea
where  $V^{(PT)}_{m_2,rat}(\theta)$ is given by Eq. (\ref{ext_theta_pt}). 
In this case, we obtain a complete solution of 
this potential by considering the same form of the $\phi$ and $r$  dependent terms (as defined in the first example) with a 
new form of $\theta$ dependent term which is now complex but PT-invariant. 

 {\bf Case (i) For $m_1=m_2=m_3=1$}

For this particular case, on comparing the above Eq. (\ref{pot_extd_pt}) with Eq. (\ref{v_rat}) (by defining $h=PT$ for $\theta$ dependent term and
$h=I$ for $\phi$ dependent term), we get the $PT$ symmetric extended potential 

\ba\label{extd_theta_pt}
V^{(PT)}_{1,ext}(\theta)= V^{PT}_{con}(\theta)+ V^{(PT)}_{1,rat}(\theta),
\ea
where 
\be\label{con_pt}
V^{(PT)}_{con}(\theta )=C\cosec^2\theta+iD\cot\theta,
\ee
is the conventional $PT$ symmetric trigonometric Eckart potential\footnote{Which is 
easily obtained by complex co-ordinate transformation $x\rightarrow ix$ of the 
rationally extended hyperbolic Eckart potential given in \cite{cq_12}.}  and the associated rational term
\ba
V^{(PT)}_{1,rat}(\theta)&=&\frac{1}{A^2(A-1)^2}\bigg[\frac{-4iB[A^2(A-1)^2-B^2]}{(iB+A(A-1)\cot\theta )}\nonumber\\
&+& \frac{2[A^2(A-1)^2-B^2]^2}{(iB+A(A-1)\cot \theta )^2}\bigg].
\ea
 The form of the $\phi$ and the $r$ dependent extended terms will be same as 
defined by Eqs. (\ref{extd_phi_1}) and (\ref{extd_rad_1}). The solution of Eq. (\ref{ucext_2}) (for $m_2=1$ and $h=PT$) with the above potential (\ref{extd_theta_pt})  is not in 
the exact form of EOPs rather they are written in the form of some types of new polynomials (discussed in detail in Ref. \cite {cq_12}) given as
\be\label{wfpt_eck_1}
\Theta^{PT}_{1,n_2}(\theta)\propto \frac {(z-1)^{\frac{\alpha_{n_2}}{2}}(z+1)^{\frac{\beta_{n_2}}{2}}}{(iB+A(A-1)\cot \theta) )}y^{(A,B)}_{n_2}(z),
\ee
with $z=i\cot \theta $. Here the polynomial function $y^{(A,B)}_{n_2}(z)$ 
can be expressed in terms of the classical Jacobi 
polynomials $P^{(\alpha_{n_2},\beta_{n_2})}_{n_2}(z)$ as
\ba
y^{(A,B)}_{n_2}(z)&=&\frac{2(n_2+\alpha_{n_2})(n_2+\beta_{n_2})}{(2n_2+\alpha_{n_2}+\beta_{n_2})}q^{(A,B)}_1(z)P^{(\alpha_{n_2},\beta_{n_2})}_{n_2-1}(z)\nonumber\\
&-&\frac{2(1+\alpha_1 ) (1+\beta_1 )}{(2+\alpha_1+\beta_1)}P^{(\alpha_{n_2},\beta_{n_2})}_{n_2}(z).
\ea
Here $q^{(A,B)}_1(z)=P^{(\alpha_1,\beta_1)}_{1}(z)$ (Classical Jacobi 
polynomial $P^{(\alpha_1,\beta_1)}_{n_2}(z)$ for $n_2=1$). The parameters $\alpha_{n_2}$ and $\beta_{n_2}$ in terms of $A$ and $B$  are 
given by
\be
\alpha_{n_2}=-(A-1+n_2)+\frac{B}{(A-1+n_2)}; \quad \beta_{n_2} =-(A-1+n_2)-\frac{B}{(A-1+n_2)}.
\ee
with
\be
A=\frac{1}{2}+\sqrt{C+m^2}, \quad \mbox{and}\quad B=\frac{D}{2}.
\ee
The other two parameters $\alpha_1$ and $\beta_1$ are simply obtained by putting $n_2=1$ in $\alpha_{n_2}$ and  $\beta_{n_2}$.
The energy eigenvalues are given by  
\be\label{enpt_1}
\ell^2=(A-1+n_2)^2+\frac{B^2}{(A-1+n_2)^2}; \quad n_2=0,1,2,...
\ee
 Thus the complete wavefunction associated with the extended $PT$ symmetric complex non-central 
potential Eq. (\ref{v_rat}) (for $m_1=m_2=m_3=1$) is obtained by using Eq. (\ref{wf_non_rat}), which is a product of the
$X_1$ Jacobi polynomial (as given by Eq. (\ref{ext_phi_sol_1})), $X_1$ Laguerre polynomial (as given by Eq. (\ref{ext_rad_sol_1})) times 
a new polynomial  given in Eq. (\ref{wfpt_eck_1}).

{\bf Case (ii) For any positive integer values of $m_1, m_2$ and $m_3$}

Again by considering the same form of the $\phi$ and $r$ dependent terms for any arbitrary values of $m_3$ and $m_1$,
 the above complex potential can be generalized  
easily for any non-zero positive integer values of $m_2$ by defining
\ba\label{extd_theta_ptm}
V^{(PT)}_{m_2,ext}(\theta)= V^{(PT)}_{con}(\theta)+ V^{(PT)}_{m_2,rat}(\theta),
\ea
where $V^{PT}_{con}(\theta )$ and $V^{(PT)}_{m_2,rat}$ are given by Eqs. (\ref{con_pt}) and (\ref{ext_theta_pt}) respectively. 
The wavefunction associated with this potential corresponding to the Eq. (\ref{ucext_2}) is given by  
\be\label{wfpt_eckm}
\Theta^{(PT)}_{m_2,n_2}(\theta)\propto \frac{(z-1)^\frac{\alpha_{n_2}}{2}(z+1)^\frac{\beta_{n_2}}{2}}{q^{(A,B)}_{m_2}(z)}y^{(A,B)}_{\nu,m_2}(z); \quad \nu=n_2+m_2-1,
\ee
where $q^{(A,B)}_{m_2}(z)=P^{(\alpha_{m_2},\beta_{m_2})}_{m_2}$ and the polynomial 
function $y^{(A,B)}_{\nu,m_2}(z)$ is 
\ba\label{y}
y^{(A,B)}_{\nu,m_2}(z)&=&\frac{2(n_2+\alpha_{n_2})(n_2+\beta_{n_2})}{(2n_2+\alpha_{n_2}+\beta_{n_2})}q^{(A,B)}_{m_2}(z)P^{(\alpha_{n_2},\beta_{n_2})}_{n_2-1}(z)\nonumber\\
&-&\frac{2(m_2+\alpha_{m_2})(m_2+\beta_{m_2})}{(2m_2+\alpha_{m_2}+\beta_{m_2})}q^{(A+1,B)}_{m_2-1}(z)P^{(\alpha_{n_2},\beta_{n_2})}_{n_2}(z),
\ea
with the parameters
 \ba
\alpha_{m_2}=-(A-1+m_2)+\frac{B}{(A-1+m_2)}; \quad  \beta_{m_2} =-(A-1+m_2)-\frac{B}{(A-1+m_2)}.\nonumber\\
\ea
The energy eigenvalues will be same as given by Eq. (\ref{enpt_1}). Thus
 the complete wavefunction and the eigenvalues of this complex non-central extended potential 
are obtained by using Eqs. (\ref{wf_non_rat}) and  (\ref{en123}).

{\footnotesize Table I.  In this table all the four forms of  
$V^{(h)}_{m_2, ext}(\theta )$ (for $h=I,II, (PT)_1$ and $(PT)_2$) with their 
corresponding energy eigenvalues ($\ell^2$) and the eigenfunctions 
($\Theta ^{(h)}_{m_2,n_2}(\theta )$) are given. Cases (i) and (iii) are
 discussed in detail in the text.}

\begin{tabular}{|l|*3{c|}} \hline
\textbf{$V^{(h)}_{m_2, ext}(\theta )$}&\pbox{30cm}{ \textbf{$\ell^2$}}& $\Theta ^{(h)}_{m_2,n_2}(\theta )$\\ \hline
\pbox{60cm} {$(i) \quad V^{(I)}_{m_2,ext} (\theta )$}
&\pbox{60cm} {$(2n_2 +\alpha +\beta +1)^2$\\
$n_2 = 0,1,2,...$\\
$\alpha =\sqrt{C+m^2}$ \\
$\beta =\frac{1}{2}\sqrt{1+4D}$ }&
\pbox{60cm} {$\frac{(1-z)^{\frac{\alpha }{2}+\frac{1}{4}}(1+z)^{\frac{\beta }{2} +\frac{1}{4}} } {P^{(-\alpha -1, \beta -1)}_{m_2} (z)}
\hat{P}^{(\alpha ,\beta )}_{n_2+m_2} (z) ;$ \\
$z = \cos 2\theta $\\
$m_2=1,2... $}\\ \hline
\pbox{60cm} {$(ii) \quad V_{m_2, ext}^{(II)}(\theta )$}
&\pbox{60cm} {$(n_2 +\frac{\alpha +\beta +1}{2})^2$\\
$\alpha =\sqrt{C+m^2 -D}$ \\
$\beta =\sqrt{C+m^2 +D}$ }&
\pbox{60cm} {$\frac{(1-z)^{\frac{\alpha }{2}+\frac{1}{4}}(1+z)^{\frac{\beta }{2} +\frac{1}{4}}} {P^{(-\alpha -1, \beta -1)}_{m_2} (z)}
\hat{P}^{(\alpha ,\beta )}_{n_2+m_2} (z) ;$ \\
$z = \cos \theta $}\\ \hline
\pbox{60cm} {$(iii) \quad V_{m_2, ext}^{(PT)_1}(\theta )$}
&\pbox{60cm} {$(A-1+n_2)^2 + \frac{B^2}{(A-1+n_2)^2}$\\
$A=\frac{1}{2} +\sqrt{C+m^2}; B =\frac{D}{2}$ \\
$ \alpha _{n_2} = -(A-1+n_2)+ \frac{B}{(A-1+n_2)}$ \\
$\beta _{n_2} = -(A-1+n_2)- \frac{B}{(A-1+n_2)} $ }&
\pbox{60cm} {$\frac{(z-1)^{\frac{\alpha _{n_2}}{2}} (z+1)^{\frac{\beta  _{n_2}}{2}}}{q_{m_2}^{(A, B)} (z)}y^{(A,B)}_{\nu,m_2}(z);$ \\
$z=i \cot \theta $ \\
$\nu  = n_2 +m_2-1,$}\\ \hline

\pbox{60cm} {$(iv) \quad V_{m_2, ext}^{(PT)_2}(\theta )$}
&\pbox{60cm} {$(n_2 +\frac{\alpha +\beta +1}{2})^2=(n_2+A)^2$\\
$\alpha =\sqrt{C+m^2 -iD}$ \\
$\beta =\sqrt{C+m^2 +iD}$ \\
$C+m^2=-B^2+(A-\frac{1}{2})^2$\\
$D=2AB$\\}&
\pbox{60cm} {$\frac{(1-z)^{\frac{\alpha }{2}+\frac{1}{4}}(1+z)^{\frac{\beta }{2} +\frac{1}{4}}} {P^{(-\alpha -1, \beta -1)}_{m_2} (z)}
\hat{P}^{(\alpha ,\beta )}_{n_2+m_2} (z) ;$ \\
$z = \cos \theta $}\\ \hline

\end{tabular} \pagebreak

{\footnotesize Table II.   The three different forms of  $\phi$ dependent terms $V^{(h)}_{m_3, ext}(\phi)$ (for $h=I,II,(PT)_1$) with their corresponding energy eigenvalues ($m^2$) and the eigenfunctions ($\Phi ^{(h)}_{m_3,n_3}(\phi)$) are given. Out of these, case (i) is already considered in detail in the text.}

\begin{tabular}{|l|*3{c|}} \hline
\textbf{$U^{(h)}_{m_3, ext}(\phi)$}&\pbox{30cm}{ \textbf{$m^2$}}& $\Phi ^{(h)}_{m_3,n_3}(\phi)$\\ \hline
\pbox{60cm} {$ (i) \quad  U^{(I)}_{m_3,ext} (\phi  )$}
&\pbox{60cm} {$p^2(2n_3 +\tilde{\alpha } +\tilde{\beta } +1)^2$\\
$n_3 = 0,1,2,...$\\
$\tilde{\alpha }  =\frac{1}{2}\sqrt{1+\frac{4G}{p^2}}$ \\
$\tilde{\beta }  =\frac{1}{2}\sqrt{1+\frac{4F}{p^2}}$ }&
\pbox{60cm} {$\frac{(1-z)^{\frac{\tilde{\alpha } }{2}+\frac{1}{4}}(1+z)^{\frac{\tilde{\beta } }{2}+\frac{1}{4}}} {P^{(-\tilde{\alpha }  -1, \tilde{\beta }  -1)}_{m_3} (z)}
\hat{P}^{(\tilde{\alpha }  ,\tilde{\beta } )}_{n_3+m_3} (z) ;$ \\
$z = \cos (2p\phi)  $ \\
$p=1,2,3,...$\\
$m_3=1,2,3....$}\\ \hline
\pbox{60cm} {$(ii) \quad U_{m_3, ext}^{(II)} (\phi )$}
&\pbox{60cm} {$p^2(n_3 +\frac{\tilde{\alpha }  +\tilde{\beta }  +1}{2})^2$\\
$\tilde{\alpha }  =\frac{1}{2} \sqrt {1+\frac{4G}{p^2} -\frac{4F}{p^2}}$ \\
$\tilde{\beta }  =\frac{1}{2} \sqrt {1+\frac{4G}{p^2} +\frac{4F}{p^2}}$ }&
\pbox{60cm} {$\frac{(1-z)^{\frac{\tilde{\alpha } }{2}+\frac{1}{4}}(1+z)^{\frac{\tilde{\beta } }{2}+\frac{1}{4}}} {P^{(-\tilde{\alpha }  -1, \tilde{\beta }  -1)}_{m_3} (z)}
\hat{P}^{(\tilde{\alpha } ,\tilde{\beta } )}_{n_3+m_3} (z) ;$ \\
$z = \cos (p\phi )$\\
$p=1,3,5,...$}\\ \hline
\pbox{60cm} {$(iii) \quad U_{m_3, ext}^{(PT)_1} (\phi )$}
&\pbox{60cm} {$p^2(n_3 +\frac{\tilde{\alpha }  +\tilde{\beta }  +1}{2})^2=(n_3p+A)^2$\\
$\tilde{\alpha }  =\frac{1}{2} \sqrt {1+\frac{4G}{p^2} -\frac{4iF}{p^2}}$ \\
$\tilde{\beta }  =\frac{1}{2} \sqrt {1+\frac{4G}{p^2} +\frac{4iF}{p^2}}$\\
$G=A^2-B^2-Ap$\\
$F=B(2A-p)$ }&
\pbox{60cm} {$\frac{(1-z)^{\frac{\tilde{\alpha } }{2}+\frac{1}{4}}(1+z)^{\frac{\tilde{\beta } }{2}+\frac{1}{4}}} {P^{(-\tilde{\alpha }  -1, \tilde{\beta }  -1)}_{m_3} (z)}
\hat{P}^{(\tilde{\alpha } ,\tilde{\beta } )}_{n_3+m_3} (z) ;$ \\
$z = \cos (p\phi )$\\
$p=1,3,5,...$}\\ \hline

\end{tabular}

\section{RE non-central potentials in $2$-Dimensions}

In two dimensional polar co-ordinates $(r,\phi)$, the Schr\"odinger equation corresponding 
to the non-central potential $V_{m_1,m_3}(r,\phi)$ is given by $(\hbar=2m=1)$ 
\be\label{sch_2d1}
\bigg[-\frac{d^2}{dr^2}-\frac{1}{r}\frac{d}{dr}-\frac{1}{r^2}
\frac{d^2}{d\phi^2}\bigg]\Psi(r,\phi)+V_{m_1,m_3}(r,\phi)\Psi(r,\phi)= E\psi(r,\Phi)\,.
\ee
The forms of the non-central potential in this co-ordinate system is given by
\be\label{pot_2d}
V^{(h)}_{m_1,m_3}(r,\phi)=\tilde{U}_{m_1,ext}(r)+\frac{1}{r^2}U^{(h)}_{m_3,ext}(\phi ),
\ee
with $h=1,2,(PT)_1$ and $(PT)_2$. 
The above Schr\"odinger equation is exactly solvable, if we define the wave function in the form
\be\label{wf_2d1}
\Psi(r,\phi)=\frac{R_{m_1}(r)}{r^{1/2}}\Phi^{(h)}_{m_3}(\phi )\,.
\ee 
Using $\Psi(r,\phi)$ in Eq. (\ref{sch_2d1}), the angular component of the wave 
function satisfies the equation
\be\label{phi_2d} 
\bigg[-\frac{d^2}{dr^2}+U_{m_3,ext}(\phi)\bigg]\Phi^{(h)}_{m_3}(\phi)=m^2\Phi^{(h)}_{m_3}(\phi),
\ee 
and the radial component satisfies
\be\label{rad_2d} 
\bigg[-\frac{d^2}{dr^2}+\tilde{U}_{m_1,ext}(r)+\frac{\big(m^2-\frac{1}{4}\big)}{r^2}\bigg]R_{m_1}(r)=ER_{m_1}(r),
\ee 
where $m^2$ is the eigenvalue of the angular equation.

These two equations (\ref{phi_2d}) and (\ref{rad_2d}) are identical to the
corresponding  
equations (\ref{ucext_1}) and (\ref{ucext_3}) 
obtained in the three dimensional case except the parameter $\ell$ in the 
three dimensional case is now replaced by $m$ in the radial 
component of the Eq. (\ref{rad_2d}). 
In this case, we have one choice of $\tilde{U}_{m_1,ext}(r)$ and four choices of $V^{(h)}_{m_3,ext}(\phi)$ (two real and two $PT$ symmetric). 
 The form of  $\tilde{U}_{m_1,ext}(r)$ will be same as in the case of three dimensions with the parameter $\tilde{\delta}=\sqrt{\delta+m^2}$.
Out of these four choices of $V^{(h)}_{m_3,ext}(\phi)$, three ( two real and one $PT$ symmetric, $h=I,II,(PT)_1$) are already discussed in Table $II$
of the previous section while the fourth form of the potential ($h=(PT)_2$) is 
special to the two dimensions.

The form of this complex potential is given by
\be\label{ext_phi_2pt}
V^{(PT)_2}_{m_3,ext}(\phi)=\frac{G}{\sin^2 (p\phi)}+\frac{iF}{\tan (p\phi)}+V^{(PT)_2}_{m_2,rat}(\phi),
\ee
where 
\ba
V^{(PT)_2}_{m_3,rat}(\phi) &=& -2p^2\cosec^2(p\phi)\bigg[ 2i\cot(p\phi) \frac{\dot{q}^{(A/p,B/p)}_{m_3}(z)}{q^{(A/p,B/p)}_{m_3}(z)}-\cosec^2(p\phi)\nonumber\\
&\times & \Bigg( \frac{\ddot{q}^{(A/p,B/p)}_{m_3}(z)}{q^{(A/p,B/p)}_{m_3}(z)}-\bigg( \frac{\dot{q}^{(A/p,B/p)}_{m_3}(z)}{q^{(A/p,B/p)}_{m_3}(z)}\bigg )^2 \Bigg )-m_3\bigg ];\quad m_3=1,2,3....\nonumber
\ea 
with $z=\cos (p\phi);\quad 0\leq p\phi\leq \pi$ and 
\be
q_{m_3}^{(A/p, B/p)}(z)=P^{(\tilde{\alpha}_{m_3},\tilde{\beta}_{m_3})}_{m_3}(z)
\ee
is a classical Jacobi polynomials with the parameters 
\ba
\tilde{\alpha} _{m_3} &=& -(A/p-1+m_3)+ \frac{B/p}{(A/p-1+m_3)} \nonumber\\
\tilde{\beta} _{m_3} &=& -(A/p-1+m_3)- \frac{B/p}{(A/p-1+m_3)}.
\ea
Here $p$ is restricted to the positive odd integers only and a dot on 
$q_{m_3}^{(A/p, B/p)}(z)$ indicates single derivative with $z$.

An explanation is in order as to why the potential as given by 
Eq. (\ref{ext_phi_2pt}) is a PT-symmetric complex potential in two
dimensions but only a complex but  not PT-symmetric in three
dimensions. The point is that unlike three space dimensions, parity in two
space dimensions correspond to say $x \rightarrow -x, y \rightarrow +y$, i.e.
it corresponds to  ($P: r\rightarrow r, \phi\rightarrow \pi-\phi$). The time 
reversal corresponds to ($T: t\rightarrow -t, i\rightarrow -i$) symmetry, the 
above potential (\ref{ext_phi_2pt}) is $PT$ symmetric in $2$-dimensions but
not in three space dimensions since in three dimensions $\phi \rightarrow
\pi + \phi$. 

Of course one can also consider the potential as given by 
Eq. (\ref{ext_phi_2pt}) in three dimensions and there it is merely complex
but non $PT$-symmetric potential. However, the spectrum is still real thereby
confirming the well known fact that $PT$-symmetry is sufficient but not 
necessary for the spectrum to be real. Note that we also have considered
this type of $\theta$-dependent potential in case $(iii)$ of Table I (detail
solution is also given in Section $2.2$), since such a potential is indeed
complex and PT-invariant term. This is because, in three dimensions, under
parity, unlike $\phi$,  $\theta \rightarrow \pi - \theta$. 
Following the same procedure as in the three dimensional case, the solution of 
this $2$-dimensional 
$PT$  symmetric non-central potential  
\be\label{v_2d}
V^{(PT)_2}_{m_1,m_3}(r,\phi)=\tilde{U}_{m_1,ext}(r)+\frac{1}{r^2}U^{(PT)_2}_{m_3,ext}(\phi),
\ee
can also be obtained in a straightforward way. In particular, the solution for 
the  special case of $m_1=m_3=1$  
is straightforward one, therefore we only consider the general case of any 
arbitrary positive integers $m_1$  and $m_3$.

Using Eq. (\ref{ext_phi_2pt}) in the angular equation
 (\ref{phi_2d}) we get the solution of the form 
\be\label{ext_phi_sol_2d}
\Phi^{(PT)_2}_{m_3,n_3}(\phi)\propto \frac{(z-1)^{\frac{\tilde{\alpha} _{n_3}}{2}} (z+1)^{\frac{\tilde{\beta}  _{n_3}}{2}}}{q_{m_3}^{(A/p, B/p)} (z)}y^{(A/p,B/p)}_{\nu,m_3}(z),
\ee
where
\ba 
 \tilde{\alpha} _{n_3} &=& -(A/p-1+n_3)+ \frac{B/p}{(A/p-1+n_3)}\nonumber\\
\tilde{\beta} _{n_3} &=& -(A/p-1+n_3)- \frac{B/p}{(A/p-1+n_3)}, 
 \ea
 and the form of  $y^{(A/p,B/p)}_{\nu,m_3}(z)$ will be same as given by Eq. (\ref{y}) (where we replace $n_2\rightarrow n_3, m_2\rightarrow m_3, A\rightarrow A/p,  B\rightarrow B/p$ and $z\rightarrow \cos(p\phi)$).
The energy eigenvalue $m^2$ is given by
\be
m^2=p^2\bigg[(A/p-1+n_3)^2 + \frac{B^2/p^2}{(A/p-1+n_3)^2}\bigg];\quad n_3=0,1,2,3...
\ee 
The parameters $A$ and $B$ in terms of $F$ and $G$ are related as 
\be
\frac{A}{p}=\frac{1}{2} +\frac{1}{2}\sqrt{1+4G/p^2}; \quad 2B =F.
\ee 
Now using equation (\ref{ext_r}) in the radial part of the Schr\"odinger equation (\ref{rad_2d}) we get 
\be\label{ext_rad_sol_2dm}
R_{m_1,n_1}(r)\propto \frac{r^{(\sqrt{\delta+m^2}+1/2)}\exp{\big(-\frac{\omega r^2}{4}\big)}}
{L_{m_1}^{(\sqrt{\delta+m^2}-1)}\big(-\frac{\omega r^2}{2}\big)}\hat{L}_{n_1+m_1}^{(\sqrt{\delta+m^2}) }\big(\frac{\omega r^2}{2}\big);\qquad 0< r < \infty,
\ee
with the energy eigenvalues 

\be\label{en2d}
E_{n_1,n_3}=\omega(2n_1+1+\tilde{\delta}), 
\ee  
where 
\be\label{del}
\tilde{\delta}=\sqrt{\delta+m^2}.
\ee
Thus the complete wavefunction and the eigenspectrum for the above seven 
parameter (four continuous and three discrete) family of potential 
(\ref{v_2d}) are  given by Eqs. (\ref{wf_2d1}) and (\ref{en2d}).
In the particular case of $m_1=m_3=0$, these potentials are reduced to their corresponding conventional potentials whose 
solutions are associated with the classical orthogonal polynomials.
 
\section{Summary}

In this work, we have constructed twelve rationally extended non-central real 
and $PT$ symmetric complex potentials in three dimensional spherical polar co-ordinates. The solutions of these 
potentials are obtained by using  
the recently discovered rationally extended potentials whose solutions are 
in terms of $X_{m_1}, X_{m_2} $ or $X_{m_3}$ exceptional Laguerre and (or) Jacobi 
orthogonal polynomials. The eigenfunctions  
and the energy eigenvalues of these twelve extended non-central potentials are 
obtained explicitly and shown that the eigenfunctions 
of these extended non-central potentials are the product of Laguerre and Jacobi EOPs. It is found that the 
three dimensional Schr\"odinger equation is exactly solvable for  the one possible choice for
$\tilde{U}_{m_1,ext}(r)$, four possible choices for $V^{(h)}_{m_2,ext}(\theta)$ 
and  three choices for $U^{(h)}_{m_3,ext}(\phi)$.
All possible choices of $\theta$ and $\phi$ dependent potentials and the 
corresponding  solutions are listed in Tables $I$ and $II$. 
 The various combinations of $\tilde{U}_{m_1,ext}(r), V^{(h)}_{m_2,ext}\quad (\theta)$ and $U^{(h)}_{m_3,ext}(\phi)$ lead to the 
total twelve different forms (four real and eight $PT$ symmetric complex) of 
the RE non-central potentials.
In the examples of $PT$ symmetric cases, some of the solutions corresponding 
to the $\theta$ 
dependent term is not in the exact form of EOPs, they are written in the forms of some types of new orthogonal polynomials ($y^{(A,B)}_{\nu,m_2}(z)$) which 
are simplified further in the terms of classical Jacobi polynomials. 
 In this works we have only consider one choice of $r$ dependent extended potential as a RE radial oscillator case. 
One can also replace the RE radial oscillator part with the conventional coulomb $U_{con}(r)=-\frac{e^2}{r}+\frac{\delta}{r^2}$ 
(as shown in Ref. \cite{kb}) then one will
have nine parameters RE non-central potential and the spectrum can also be obtained easily. Few attempts at rational 
extension of Coulomb have been done \cite{yg2}, but they are not very general, so we are not mentioning them.

In a particular case of $m_1=m_2=m_3=0$, these potentials are reduced to their 
conventional counterparts (which are non-rational) with seven parameters (six continuous and one discrete) 
whose solutions are in terms of classical orthogonal polynomials.
Out of these twelve conventional cases, the eight $PT$ symmetric complex non-central seven parameters conventional potentials 
are also new and not discussed earlier.   

We have also considered the Schr\"odinger equation in two dimensional 
polar co-ordinates and constructed four possible forms (two real 
and two $PT$ symmetric complex) of the seven parameters (four continuous and 
three discrete) RE non-central potentials. The solutions of these potentials 
are also obtained in terms of EOPs.

{\bf Acknowledgments}
 B.P.M. acknowledges the financial support 
from the Department of Science and Technology (DST), Gov. of India under SERC project sanction
grant No. $SR/S2/HEP-0009/2012$. A.K. wishes to thank Indian National Science Academy (INSA) for 
the award of INSA senior scientist position at Savitribai Phule Pune University.

\end{document}